\documentclass[grl]{agutex}

\usepackage{lineno}
\linenumbers*[1]

  \usepackage[dvips]{graphicx}
  \setkeys{Gin}{draft=false}

\authorrunninghead{KRIVOVA ET AL.}

\titlerunninghead{ACRIM-GAP AND TSI TREND REVISITED}

\authoraddr{N. A. Krivova,
Max Planck Institute for Solar System Research,
Max-Planck-Str. 2, 37191 Katlenburg-Lindau, Germany
(natalie@mps.mpg.de)}

\authoraddr{S. K. Solanki,
Max Planck Institute for Solar System Research,
Max-Planck-Str. 2, 37191 Katlenburg-Lindau, Germany}

\authoraddr{T. Wenzler,
Hochschule f\"ur Technik Z\"urich, CH-8004 Z\"urich,
Switzerland}

\begin{document}

\title{ACRIM-gap and total solar irradiance revisited:\\
Is there a secular trend between 1986 and 1996?}

\authors{N. A. Krivova, \altaffilmark{1}
S. K. Solanki, \altaffilmark{1,}\altaffilmark{2}
and T. Wenzler \altaffilmark{1,}\altaffilmark{3}}

\altaffiltext{1}{Max Planck Institute for Solar System Research,
D-37191 Katlenburg-Lindau, Germany}

\altaffiltext{2}{School of Space Research, Kyung Hee University,
Yongin, Gyeonggi 446-701, Korea}

\altaffiltext{3}{Hochschule f\"ur Technik Z\"urich, CH-8004 Z\"urich,
Switzerland}

\begin{abstract}

A gap in the total solar irradiance (TSI) measurements between ACRIM-1 and
ACRIM-2 led to the ongoing debate on the presence or not of a secular trend
between the minima preceding cycles 22 (in 1986) and 23 (1996). It was
recently proposed to use the SATIRE model of solar irradiance variations to
bridge this gap. When doing this, it is important to use the appropriate
SATIRE-based reconstruction, which we do here, employing a reconstruction
based on magnetograms. The accuracy of this model on months to years
timescales is significantly higher than that of a model developed for
long-term reconstructions used by the ACRIM team for such an analysis.
The constructed `mixed' ACRIM~--- SATIRE composite shows no increase in the
TSI from 1986 to 1996, in contrast to the ACRIM TSI composite.


\end{abstract}

\begin{article}

\section{Introduction}
\label{intro}

Solar total irradiance has been measured by a number of space-borne
instruments without interruptions since 1978 \citep{froehlich-2006}.
However, no single instrument remained operational over the whole period,
and a cross-calibration and construction of a composite of measurements by
several radiometers is not trouble-free.
A detailed description of individual data sets, their problems and
corrections is given by \citet{froehlich-2006}, cf.
\citet{scafetta-willson-2009}.
As a consequence, three different composites of the total solar irradiance
(TSI) have been constructed: PMOD \citep{froehlich-2006}, ACRIM
\citep{willson-mordvinov-2003}, and IRMB \citep{dewitte-et-al-2004}.

The main disagreement concerns the TSI levels during the minima preceding
cycles 22 (which took place in 1986) and 23 (in 1996) and thus the presence
of a long-term trend during this period.
The origin of the discrepancy lies in the fact that the launch of the
ACRIM-2 experiment on UARS, originally planned to overlap with ACRIM-1 on
SMM, had to be postponed.
Thus for more than 2 years (July 1989 to October 1991; the so-called ACRIM
gap) only the data from HF/Nimbus7 and ERBS/ERBSE are available.
The ERBS irradiance measurements were made only approximately every 2 weeks.
In September 1989, the HF radiometer was switched off for several days due
to saturation of its output signal.
When it returned to normal operating mode, it apparently showed an abrupt
increase in irradiance by about 0.4~W\,m$^{-2}$ \citep[for more details,
see][]{lee-et-al-1995,chapman-et-al-96,lockwood-froehlich-2008}.
There is also an indication for another either abrupt or gradual increase of
a similar magnitude \citep{lee-et-al-1995,froehlich-2006}.
This change in the HF sensitivity was allowed for in the PMOD composite but
not in the ACRIM one, which led to the disagreement in the long-term
behaviour mentioned above.
The ACRIM composite shows an increase in the TSI of 0.037\% (about
0.5~W\,m$^{-2}$) between the minima of 1986 and 1996
\citep{willson-mordvinov-2003}, whereas the PMOD composite gives nearly the
same TSI values (difference less than 0.06~W\,m$^{-2}$ or 0.004\%) for both.

Recently, \citet{scafetta-willson-2009} have proposed to use the SATIRE
model of TSI variations, in order to bridge the ACRIM gap and thus estimate
the long term change independently.
The SATIRE models
\citep{solanki-et-al-2005a,krivova-solanki-2008a}
calculate variations of solar irradiance from the solar surface magnetic
field evolution.
Several versions of the model have been developed for different
applications.
Each version is optimised for a different time scale.
For reconstructions restricted to the period after 1974,
direct full-disc
measurements of the solar photospheric magnetic field, i.e. magnetograms,
can be employed together with continuum images to identify magnetic features
(i.e. sunspots, faculae and the network)
\citep{krivova-et-al-2003a,wenzler-et-al-2004a,wenzler-et-al-2005a,%
wenzler-et-al-2006a}.
The brightnesses of the individual types of features are time-independent
and are calculated from semi-empirical model atmospheres
\citep{unruh-et-al-99}.
Variations of TSI result from the emergence, evolution and decay of magnetic
features, which is obtained from the magnetograms and continuum images.
Thus variability on all considered time scales is modelled in a
self-consistent way, with no additional assumptions on a long-term trend.
In the following we refer to this version of the model as SATIRE-S, where
the S stands for `Satellite era'.
This model is indeed
well suited for a self-contained test of the irradiance trends around the
ACRIM gap.

In contrast, the model
SATIRE-T (where the T stands for Telescope era)
by \citet{krivova-et-al-2007a} that has been employed
by \citet{scafetta-willson-2009} is not suited for such an analysis.
It has been developed in order to provide an insight into irradiance changes
on decadal to centennial time-scales and is, by design, significantly
less accurate, particularly on time scales of months.
The main challenge of the longer-term models is that the only direct proxy
of solar activity going back to the Maunder minimum is the sunspot number.
The evolution of the bright magnetic elements (faculae, network) has to be
derived somehow from the sunspot record.
For this, \citet{krivova-et-al-2007a} use the coarse physical model developed
by \citet{solanki-et-al-2000,solanki-et-al-2002}, which allows a
reconstruction of the solar photospheric magnetic field from the sunspot
number.
Unfortunately, the evolution of the facular and network components cannot be
recovered on a daily basis using such a model: only averages over several
months can be considered.
This is clear from the Eqs.~(1--7) of that work and the corresponding values
of the involved time scales for the decay and flux transfer in active and
ephemeral regions (about 3~months to years).
Note that variations of the irradiance on time scales of days to weeks
is still well reproduced by this model, since
they are driven by the evolution of sunspots, which are adequately
represented by the daily sunspot number record.

Thus, by its very conception, the model by \citet{krivova-et-al-2007a}
employed by \citet{scafetta-willson-2009} is expected to be relatively
accurate on time scales of days to the solar rotation and from the
solar cycle to centuries, whereas its accuracy on time scales of several
months to years is limited.
This is exactly opposite to what \citet{scafetta-willson-2009} assumed.
Here we repeat their analysis employing the more accurate SATIRE-S model
by \citet{wenzler-et-al-2006a,wenzler-et-al-2009a} based on quasi-daily
full-disc magnetograms and continuum images of the Sun recorded by the
National Solar Observatory, Kitt Peak (NSO/KP) between 1974 and 2003.
We show that employing 
a more appropriate model gives a rather different
result than obtained by  \citet{scafetta-willson-2009}.
We stress that it is not the aim of this paper to produce a new TSI
composite, but rather to show that the approach taken by
\citet{scafetta-willson-2009} is completely consistent with a deeper minimum
preceding cycle 23 than that preceding cycle 22, if an appropriate model is 
used.

\section{Combining ACRIM with SATIRE-S}
\label{sect2}

We employ exactly the same technique as used by
\citet{scafetta-willson-2009}.
Thus we compare the measurements by the ACRIM-1 and ACRIM-2 instruments
directly with the SATIRE-S model, in order to construct a `mixed' composite
based on ACRIM data when they are available and the model to bridge the gap.

ACRIM-1 and 2 data are taken from the ACRIM web page: http://www.acrim.com/.
ACRIM-1 data are available between February 1980 and July 1989, ACRIM-2
between October 1991 and November 2001.

We employ the SATIRE-S model based on NSO/KP magnetograms and continuum
images as described by \citet{wenzler-et-al-2006a,wenzler-et-al-2009a}.
These data are available between 1974 and 2003.
There is a single free parameter in the model which can be adjusted to
achieve best agreement with observations.
This parameter
takes into account the saturation of brightness in regions with
concentrations of magnetic elements higher than a threshold
[see \citeauthor{wenzler-et-al-2006a},
\citeyear{wenzler-et-al-2006a}, for further details].
Note that this single parameter controls the magnitude of the variations on
all time scales under consideration and thus cannot be changed arbitrarily,
in order to best reproduce variations on a selected time scale.
\citet[][WSK09]{wenzler-et-al-2009a} ran the reconstruction with different
values of the free parameter, in order to find
the best possible fit to each of the three existing TSI composites: PMOD
(this reconstruction is referred to as WSK09 PMOD), ACRIM (WSK09 ACRIM) or
IRMB (WSK09 IRMB).
Here we use these three models as they were derived in that paper, i.e. the
free parameter is fixed to the values deduced by
\citet{wenzler-et-al-2006a} for each of the composites.

 \begin{figure*}
 \noindent\includegraphics[width=39pc]{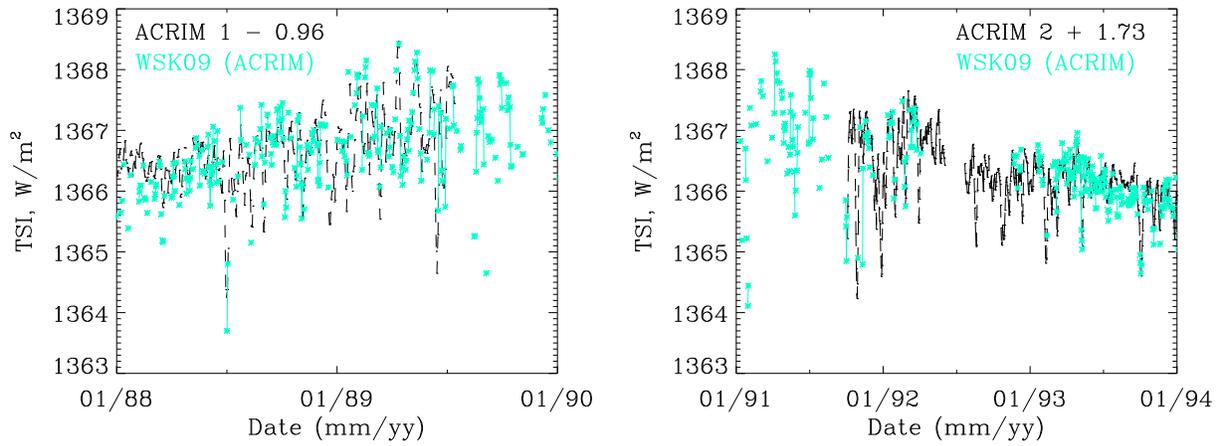}
 \caption{Measured and reconstructed TSI during 1988--1990 and
 1991--1994.
 ACRIM-1 (left panel) and ACRIM-2 (right panel) measurements (black dashed line) are
 shifted by $-0.96$~W\,m$^{-2}$ and 1.73~W\,m$^{-2}$, respectively, to the
 level of the WSK09 ACRIM model (asterisks connected by grey solid line
 when there are no gaps).
 }
 \label{fig_acrim}
 \end{figure*}

 \begin{figure*}
 \noindent\includegraphics[width=39pc]{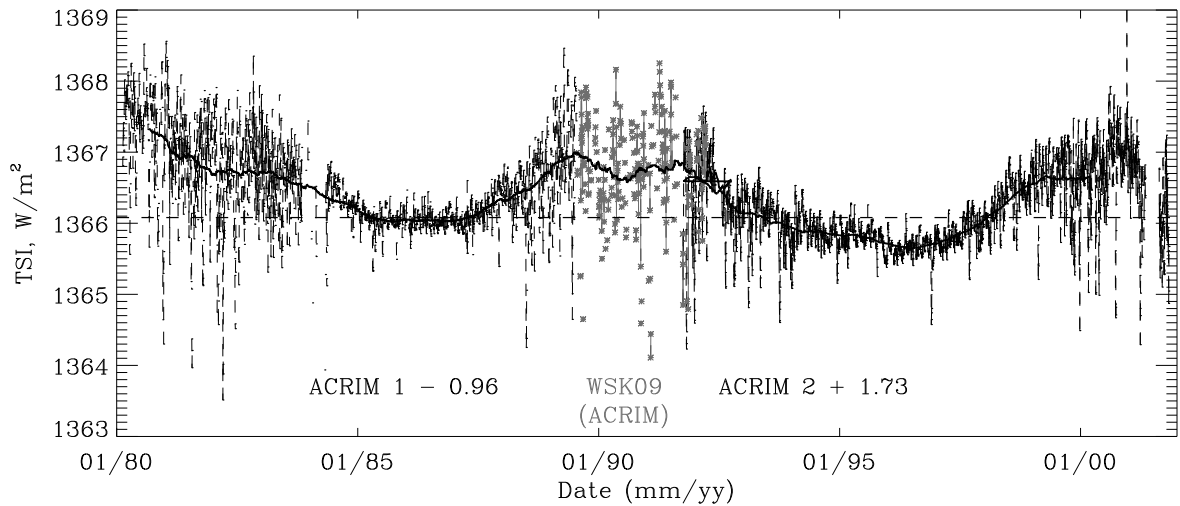}
 \caption{`Mixed' TSI composite constructed from ACRIM-1 and ACRIM-2 data
(black dashed line), with the gap bridged using the WSK09 ACRIM model
(asterisks connected by grey solid line when there are no gaps).
The heavy solid line is the 1-year smoothed TSI, and the 
horizontal dashed
line shows the level of the minimum preceding cycle 22.
}
 \label{fig_comp_acrim}
 \end{figure*}

 \begin{figure*}
 \noindent\includegraphics[width=39pc]{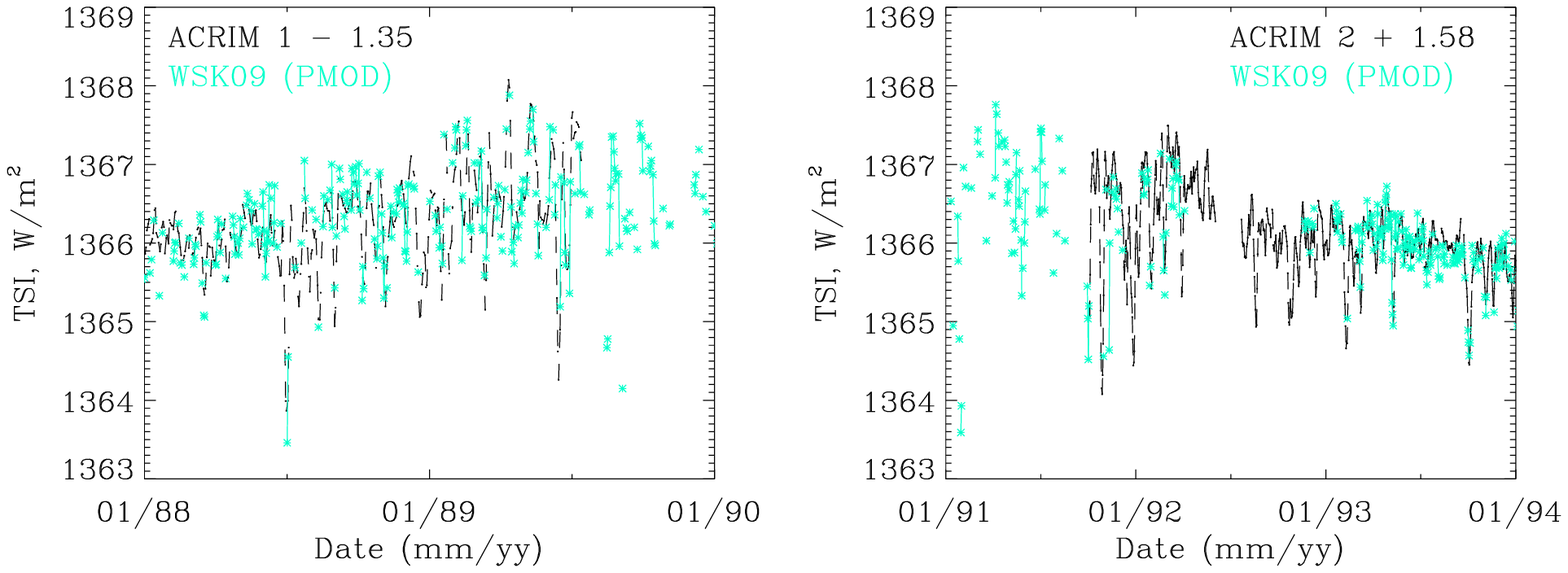}
 \caption{Measured and reconstructed TSI during 1988--1990 and
 1991--1994.
 ACRIM-1 (left panel) and ACRIM-2 (right panel) measurements (black dashed line) are
 shifted by $-1.35$~W\,m$^{-2}$ and 1.58~W\,m$^{-2}$, respectively, to the
 level of the WSK09 PMOD model (asterisks connected by grey solid line
 when there are no gaps).
 }
 \label{fig_pmod}
 \end{figure*}

 \begin{figure*}
 \noindent\includegraphics[width=39pc]{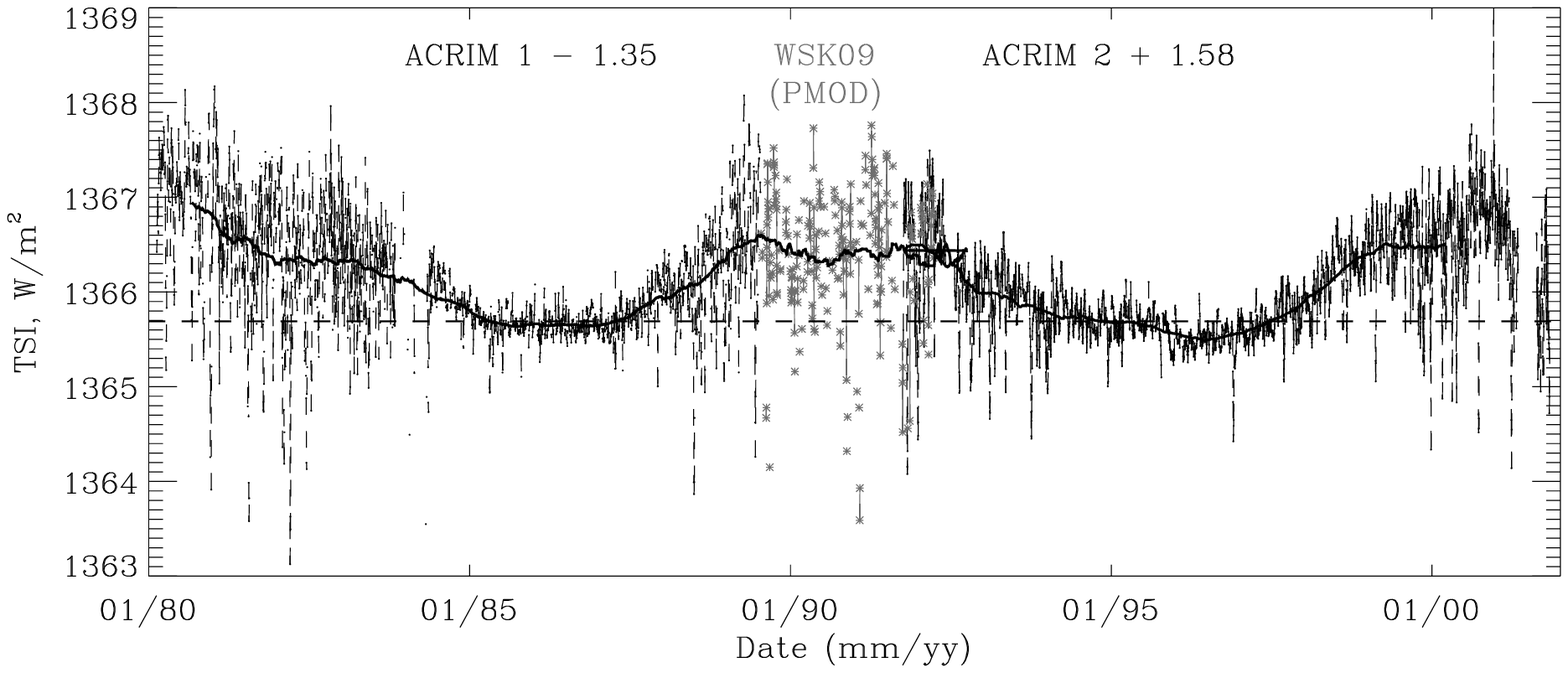}
 \caption{Same as in Fig.~\ref{fig_comp_acrim} but using the WSK09 PMOD
 model.
}
 \label{fig_comp_pmod}
 \end{figure*}

In order to compare the model with the data and produce an
ACRIM-1~--WSK09~--ACRIM-2
`mixed' composite as proposed by \citet{scafetta-willson-2009}, we use
1~year periods of overlap between the model and ACRIM-1 (July 1988~-- July
1989) or ACRIM-2 (October 1992~-- October 1993).
The periods were selected to be exactly the same as in the paper by
\citet{scafetta-willson-2009}, to avoid bias and to
allow a direct comparison to their results.
For these periods, we minimise the squared differences between the ACRIM
measurements and the SATIRE-S results, in order to
derive the shifts between their absolute levels.
For WSK09 ACRIM, this comparison is shown in Fig.~\ref{fig_acrim} and
reveals a good agreement between the model and the data.
Figure~\ref{fig_comp_acrim} shows a `mixed' composite, based on ACRIM data
prior to July 1989 and after October 1991 and on the WSK09 ACRIM model
during the ACRIM gap.
The minimum preceding cycle~23 is 0.38~W\,m$^{-2}$, or 0.028\%, lower
compared to the minimum preceding cycle~22 (as derived from the minima of
the 1-yr smoothed record).

 \begin{table}
 \caption{Summary of the comparison of the SATIRE WSK09 ACRIM, PMOD
and IRMB models with ACRIM measurements.
Listed are: model versions; derived shifts between ACRIM-1 or ACRIM-2 data and the
model for the analysed periods of overlaps; difference in the 1-yr smoothed
minimum TSI at the minimum of cycle~23 compared to the minimum of cycle~22;
the relative difference in \% normalised to the cycle~22 minimum.
}
 \begin{tabular}{ccccc}
\tableline
Model     & shift   & shift    & Change  & Relat. change\\
          & ACRIM-1 & ACRIM-2  & 23--22  & 23/22--1    \\
          &         &          & W/m$^2$ & \% \\
\tableline
WSK09 ACRIM & $-0.96$  & 1.73  & $-0.38$ & $-0.028$ \\
WSK09 PMOD  & $-1.35$  & 1.58  & $-0.15$ & $-0.011$ \\
WSK09 IRMB  & $-0.75$  & 1.95  & $-0.38$ & $-0.028$ \\
\tableline
 \end{tabular}
\label{tab:one}

 \label{table}
 \end{table}

In order to test whether this result is solid and is independent of the
value of SATIRE's free parameter, we have repeated the same procedure
with the WSK09 model optimised to best fit the PMOD
and IRMB composites, i.e. the same model,
but with different values of the free parameter
\citep[see][]{wenzler-et-al-2009a}.
The results for the PMOD composite
are shown in Figs.~\ref{fig_pmod} and \ref{fig_comp_pmod}.
For this model, the TSI at the minimum preceding cycle~23 has a value that
is still lower, by 0.15~W\,m$^{-2}$ or 0.011\%, than the TSI in the minimum
preceding cycle~22.
The difference between the two minima derived using the WSK09 IRMB model is
same as for the WSK09 ACRIM version.
All results are summarised in Table~\ref{table}.
As a further test, we have also varied the length (between 1 and 3 years)
and the dates (1 year within the periods July 1987~-- July 1989 and 
October 1991~-- October 1993) of the overlap period.
In all cases, the mean TSI during the minimum of cycle~23 is lower than that
for cycle~22 by about 0.15 to 0.7~W\,m$^{-2}$ (0.011--0.05\%),
in contrast to the result of \citet{scafetta-willson-2009} who found a
TSI increase of 0.033\% (0.45~W\,m$^{-2}$).

\section{Conclusion}

We have compared the SATIRE-S model of the TSI variations with the
measurements by the ACRIM-1 and ACRIM-2 experiments, in order to bridge the
so-called ACRIM gap (July 1989 to October 1991), as proposed by
\citet{scafetta-willson-2009}.
This gap is the source of the on-going debate about the presence of the
secular variation in solar irradiance between the minima of cycle~22 and 23.
The SATIRE-S model calculates the TSI variations from the continuously
evolving distribution of the solar surface magnetic field
obtained from NSO/KP magnetograms and continuum images
\citep{wenzler-et-al-2009a} covering the period 1974--2002.
In contrast to the SATIRE-T
model by \citet{krivova-et-al-2007a} employed by
\citet{scafetta-willson-2009}, in this model variations on all covered time
scales are modelled in a self-consistent way, with no additional assumptions
regarding the long term trend.
Also the accuracy of the SATIRE-S model is significantly higher than that of
the SATIRE-T model since it uses direct measurements of the solar
photospheric magnetic flux rather than its modelled evolution.
Thus it is best suited for such a test.

The constructed `mixed' ACRIM-1~-- WSK09~-- ACRIM-2
composite does not show an
increase in the TSI from 1986 to 1996,
in contrast to the ACRIM composite.
Independently
of the value of the model's free parameter, a
slight decrease is found.
The magnitude of this decrease cannot be estimated very accurately from such
an analysis (and therefore such a `mixed' composite should not be considered
as a replacement of real measurements), but it lies between approximately
0.15 and 0.7~W\,m$^{-2}$ (0.011--0.05\%) for different values of the model's
single free parameter.
Note that irradiance changes due to non-magnetic effects, if any,
cannot be revealed by either SATIRE-S used here nor by SATIRE-T employed by 
\citet{scafetta-willson-2009}.




\begin{acknowledgments}
This work was supported by the Deutsche Forschungsgemeinschaft, DFG project
number SO~711/2 and by the WCU grant No.~R31-10016 funded by the Korean
Ministry of Education, Science and Technology.
\end{acknowledgments}

%
%
%
%
%
%
%
%
%
%


\end{article}




%

%
%
%

\end{document}